\documentclass[]{interact}
\makeatletter\def\@received{}\makeatother
\usepackage{epstopdf}
\usepackage[caption=false]{subfig}

\usepackage{booktabs}
\usepackage{longtable}
\usepackage{array}
\usepackage{tabularx}
\usepackage{url}
\usepackage[natbibapa,nodoi]{apacite}
\makeatletter
\def\LT@makecaption#1#2#3{%
  \LT@mcol\LT@cols c{\hbox to\z@{\hss\parbox[t]\LTcapwidth{%
    \sbox\@tempboxa{#1{#2}\ #3}%
    \ifdim\wd\@tempboxa>\hsize
      #1{#2}\ #3%
    \else
      \hbox to\hsize{\hfil\box\@tempboxa\hfil}%
    \fi
    \endgraf\vskip\baselineskip}%
  \hss}}}
\makeatother

\theoremstyle{plain}

\theoremstyle{definition}

\theoremstyle{remark}

\begin{document}


\title{A Hybrid Insider Threat Detection Framework Combining Multi-Agent Simulation, Layered SIEM Correlation, and Theory-of-Mind Reasoning}

\author{
\name{Firdous Kausar, Asmah Muallem, Naw Safrin Sattar, and Mohamed Zakaria Kurdi}
\affil{Department of Computer Science and Data Science, SACS, Meharry Medical College, Nashville, USA. }
}

\maketitle

\begin{abstract}
This paper presents a hybrid insider threat detection framework for enterprise environments, integrating multi-agent simulation, layered SIEM correlation, trust-adaptive thresholds, behavioral and communication forensics, and Theory-of-Mind reasoning. Email is treated not as a control channel but as a coordination and social-engineering evidence stream correlated with authentication, file-access, and privilege events. Four variants are evaluated: Layered SIEM-Core (LSC), Cognitive-Enriched SIEM (CE-SIEM), Evidence-Gated SIEM (EG-SIEM), and EG-SIEM-Enron with Enron-calibrated email forensics. Across ten matched runs with eight malicious insiders, actor-level F1 improves from 0.567 for LSC to 0.774 for CE-SIEM, 0.898 for EG-SIEM, and 0.944 for EG-SIEM-Enron; paired Wilcoxon tests confirm the first three differences after Holm-Bonferroni correction. Evidence gating reduces confirmed false positives from 33.7 per run under LSC and 49.3 under CE-SIEM to 0.2 and 0.0 respectively, a precision gain traded against longer confirmation time. Domain-shift evaluation shows that an Enron-trained email classifier does not transfer to a different operational email domain, although target-domain fine-tuning reaches F1 = 0.974 under grouped template-family evaluation. On CERT r4.2, the evidence-gated logic improves actor-level detection while reducing false positives, outperforming tabular anomaly detectors and a sentence-embedding baseline. Scalability tests to 1000 agents indicate stable detection quality.
\end{abstract}



\section{Introduction}

Insider threats are malicious or negligent actions by authorized users that pose severe challenges to organizational security. Unlike external attacks, insiders have legitimate access, making anomalous behavior hard to distinguish. Traditional detection e.g., static signatures or one-shot ML models often yield high false positives or fail under noise. Recent work has explored machine learning and anomaly-based SIEMs, but these alone neglect human factors like intent, trust, and communication context. To address this, we propose a hybrid system that bridges agent-based reasoning, behavioral analytics, and adaptive learning. Our approach builds a synthetic MAS environment of users (benign and various attacker types) interacting over resources and messaging. A central SIEM ingests the user logs and provides successive layers of analysis. Crucially, we augment agents with a Theory-of-Mind (ToM) layer to infer intentions, and we apply behavioral forensics on communications. An online learning component continuously updates feature weights, while trust calibration adjusts alert thresholds for each user.

Enterprise and organizational IT environments exacerbate these difficulties since services depend on tightly integrated systems, for example, identity and access management, administrative interfaces, and security monitoring. Privileged operations are carried out on a control plane of remote sessions, configuration writes, and administrative actions. The difficulty for defenders is to reason over varied telemetry spanning IT and operational activity within the Security Operations Center (SOC). This includes authentication activities, for instance, SSO and VPN, endpoint and file access, role and privilege activities, and administrative control-plane logs. Simultaneously, day-to-day operations require coordination of workflow communications, for instance, using email and ticketing systems, to request changes, approve vendor activities, distribute configuration items, and request remote support. This implies that, although email is not a direct control channel for critical systems, it is a high-value evidence flow and an attack-enabler flow, which can be utilized prior to, during, and after privileged activities, for instance, social engineering to obtain credentials and access, disclosure of sensitive documentation, and malicious change requests. This creates a wide insider surface and a tremendous "atypical-but-benign" volume, making both false positives and false negatives costly in terms of operational impact.

\subsection{Operational Context and Threat Model}

Our focus is on enterprise and organizational IT environments, where services depend on the integration of authentication systems, file and data repositories, privileged administrative interfaces, and operational communication channels. The assets we are concerned with include sensitive data stores, system and configuration backups, access-control and identity systems, and the servers and networks that connect them. Operations involve sensitive assets such as system configurations, device and asset inventories, configuration backups, and vendor maintenance materials, and remote access is often provided to third-party vendors for installation, diagnostics, and support.

In this paper, an insider is described as a legitimate user (or a compromised account that is indistinguishable from a legitimate user at the log level) that has access to legitimate resources and communication channels. The defender is the Security Operations Center (SOC) that monitors heterogeneous telemetry across the IT environment, including authentication events (SSO/VPN), endpoint and file-access events, role and privilege changes, administrative and configuration-control events (such as configuration writes, privileged overrides, administrative actions, and alert acknowledgements), as well as communication metadata and content events (email). Email is not treated as a direct control or actuation channel for critical systems. Privileged operations are performed through administrative interfaces using remote sessions, configuration writes, and similar control actions. Email is used as a workflow coordination and social engineering mechanism that can occur before or concurrently with these control actions. Hence, email-based forensics is related to authentication, privileges, file access, and administrative control actions.

\begin{table}[t]
\caption{Evaluated insider scenarios, affected assets, and observable evidence streams.}
\label{tab:mapping}
\scriptsize
\centering
\begin{tabular}{p{1.9cm} p{6.2cm} p{4.9cm}}
\hline
\textbf{Scenario} & \textbf{Description and impacted assets} & \textbf{Primary evidence streams} \\
\hline
Exfiltration & Bulk export of sensitive configurations, system documentation, device inventories, or access-control records, potentially alongside procurement/HR records, from data repositories and documentation stores & File access, large downloads, after-hours access, data-access patterns, email indicators \\
Stealth & Small, incremental configuration or threshold changes over days to mimic routine activity and avoid spikes, affecting system configurations and monitoring telemetry & Low-amplitude deviations over time, repeated access patterns, configuration reads/writes, telemetry drift \\
Takeover & Privilege escalation to administrator followed by an unsafe configuration push, override, or alert suppression on the administrative control plane (IAM and systems) & Role/privilege changes, unusual admin actions, configuration activation events, session anomalies \\
Staging exfiltration & Staging compressed backups or configuration bundles on shared drives, file servers, gateways, or system backups prior to outbound transfer & Compression/archiving events, new staging directories, unusual storage growth, outbound attempts \\
Email leakage & Vendor spearphish or ``official'' work-order/change-request email used to obtain credentials/remote access or distribute malicious configuration artifacts through corporate email, collaboration tools, and the vendor ecosystem & Authorship/style drift, abnormal recipients/domains, attachment anomalies, phishing likelihood signals \\
\hline
\end{tabular}
\end{table}

Table~\ref{tab:mapping} identifies the five analyzed insider scenarios, the assets that are affected, and the main sources of evidence available to a SOC/SIEM, while noting that email is a coordination and social engineering mechanism that may precede or coexist with privileged control actions, rather than acting as a control mechanism itself.

Our key contributions are as follows.
\begin{itemize}
    \item  We propose an adaptive insider threat detection framework for enterprise and organizational settings that leverages multi-agent simulation (MAS), layered SIEM correlation, and cognitive/behavioral modeling. Our simulation models a realistic operational process where staff and vendors interact with organizational systems, shared resources, and communications.

    \item We propose a layered SIEM solution that improves detection capabilities through a stepwise process involving policy constraints, behavioral anomalies, and evidence-gated confirmation in a setting where operational noise is present.

    \item We incorporate Theory-of-Mind (ToM) intent signals and behavioral/communication forensics to provide context beyond raw logs. In particular, email is treated as an operational coordination and social-engineering channel (e.g., vendor change requests, work-order approvals, and phishing for remote access) that can be correlated with authentication anomalies and privileged control actions.

     \item Empirical evaluations are carried out for four different variants of systems: LSC, CE-SIEM, EG-SIEM, and EG-SIEM-Enron. Several runs are made for different malicious behaviors. Improved actor-level detection is achieved with a reduction in false positives when compared to matched workload conditions.

\end{itemize}

The remainder of the paper is organized as follows: Section \ref{sec:StateOfTheArt} reviews related work on insider-threat detection; Section  \ref{sec:dataset} summarizes the dataset resources and preprocessing used to support email-style and phishing-related signals; Section \ref{sec:method}  presents our proposed hybrid framework; Section \ref{sec:results}  describes the implementation and experimental evaluation, reporting results; Section \ref{sec:limit} discusses the limitations; finally, Section \ref{sec:concl}  concludes with key findings and directions for future work.

\section{Related Work}
\label{sec:StateOfTheArt}

A prominent challenge in technology is the evolving landscape of cyber attacks and their catastrophic effects on various domains. Some of the persistent attacks include malware and ransomware, phishing, social engineering, among many more. The use of any of these in anticipation of corrupting a network can be part of the act of an individual with malicious intent. Insider threat is a common and major challenge in cybersecurity affecting companies, organizations, academia, and government agencies. It has been reported that insider threats account for a fourth of all cyberattacks experienced by U.S. organizations \citep{IDG2018Cybercrime}. These types of threats are carried out by an authorized personal or employee within an organization, whom may be familiar with the structure, valued properties, and security layers of the organization \citep{le2021anomaly}. Common threats include but are not limited to disclosure of classified information, theft of personal information, trade secrets/intellectual property theft, and IT system sabotage \citep{collins2016insider}.

The use of unsupervised learning methods have been known to be useful for anomaly detection of insider threats. \citet{le2021anomaly} propose an anomaly detection approach using unsupervised learning for insider threat detection by employing four unsupervised machine learning methods with different working principles and through the exploration of various representations of data with temporal information. In their work, they apply Autoencoder and Isolation Forest models for insider threat detection and analyze their relationship with investigation budgets, and they further demonstrate that voting-based ensembles of anomaly detection methods can enhance detection performance and robustness. Another form of insider threat detection using machine learning has been employed by \citet{manju2025insider} for user behavior analysis. In their study, they employ anomaly detection using Isolation Forest and one-class SVM. Their results indicate that the Isolation Forest achieved higher recall and precision scores (0.80 and 0.92, respectively) compared to the one-class SVM (0.75 and 0.80), suggesting that the one-class SVM may have failed to identify some true insider threats and that the Isolation Forest model more accurately detected positive instances. With the accuracy being the same for both models, 0.90, but the Isolation Forest providing a higher precision result, this demonstrates that the model can excel with a false-positive associated cost \citep{manju2025insider}.

Multiagent systems approaches have been proposed for insider threat detection. To improve existing insider threat detection methods, agent-based models have been applied to monitor user activity for more effective threat identification. \citet{ali2008towards} propose an Agent-based User-Profiling model that monitors behavior of authorized users in an organization to avoid the risk. In their study, the profile of all insiders are built and maintained by the proposed model and the profile is dynamic, continuously updated while monitoring the behavior of an insider. \citet{nikolov2020evaluation} propose a multi-agent ranking framework for advanced persistent threat detection through the use of behavioral-analysis and pattern recognition. In their work, they simulate various attacks and inject them into real-world log files to emulate typical network activity, including background traffic that occurs daily and that adversaries may exploit to conceal their actions. The results demonstrate that with some exceptions, 100\% of all attacks have been ranked in the top 20, indicating an analyst would need to analyze 20 entries to capture all possible attacks in these scenarios.

Text mining enables cognitive profiling of email senders by extracting demographic and thematic markers from linguistic patterns, creating behavioral fingerprints for anomaly detection \citep{Kurdi2019}. This builds on a substantial body of work in author attribution across linguistics, psychology, and NLP, where stylistic differences reliably signal identity and intent. Representative results include 80\% gender-classification accuracy from n-grams and function words \citep{Argamon2003}, 76.8\% authorship accuracy from frequent word sequences \citep{Coyotl2006}, and 61--65\% age-range prediction on large health-forum corpora \citep{Shrestha2016}, alongside evidence that gendered language use differs systematically across large text samples \citep{Newman2008} and that character-level models support topic-neutral attribution \citep{Sarawgi2011}. We draw on this lineage only as motivation for style-deviation and authorship-consistency features; the experiments reported here do not perform demographic profiling.

More recent work applies transformer and large language model representations to
insider-threat and log anomaly detection. ITDBERT encodes temporal-semantic structure
in user activity for insider-threat classification \citep{Huang2021ITDBERT}, and Log2vec
constructs heterogeneous graph embeddings over enterprise logs to detect malicious
activity within an organization \citep{Liu2019Log2vec}. Graph-based detection has been
extended with dual-domain graph convolutional networks for adaptive insider-threat
detection \citep{Li2023DualDomain}, while statistical and sequential analysis frameworks
target the same benchmark family \citep{Xiao2024Unveiling}. On the LLM side, LogGPT
treats log sequences as natural language for anomaly detection \citep{Qi2023LogGPT},
Audit-LLM assigns specialized roles to cooperating LLM agents for log-based
insider-threat detection \citep{Song2024AuditLLM}, and RedChronos reports a
production-scale LLM log analysis system for enterprise SOCs \citep{Li2025RedChronos}.
\citet{Song2025Confront} fine-tune an LLM on natural-language renderings of behavior
logs and report strong results on CERT v6.2; that evaluation uses a different dataset
release and evaluation unit from the CERT r4.2 user-day and actor-level protocol adopted
here, so the reported scores are not directly comparable with ours. Federated
parameter-efficient tuning has been proposed to address the privacy constraints such
deployments face \citep{Wang2024FedITD}, and benchmarking studies caution that LLM
performance on log analysis varies substantially with task framing and prompt design
\citep{Karlsen2024Benchmarking}.

These approaches differ from ours in two respects. First, they operate on observed log
streams alone, whereas our framework couples log evidence with agent-internal intent
signals produced by the simulation. Second, they optimize detection at the record or
sequence level, whereas our evidence gate targets actor-level precision under an explicit
multi-evidence confirmation requirement. To quantify what a generic learned
representation contributes on tabular behavioral features, we include a sentence-encoder
embedding baseline in the CERT evaluation reported in Section~\ref{sec:cert}.

What remains largely unaddressed across these lines of work is the combination of agent-internal intent modeling with anomaly-based correlation inside a single detection pipeline. The multi-agent LLM systems described above coordinate analysis agents over observed log streams, whereas the agents in our simulation carry beliefs, goals, and plans that themselves generate the intent features the SIEM consumes. We therefore propose a hybrid framework that integrates agent-based reasoning, behavioral analytics, and adaptive learning. Adaptive learning improves existing threat detection capabilities, both in classical machine learning approaches and in proposed multi-agent detection systems, by adjusting to the evolving behaviors of malicious actors, which change gradually and strategically. The integration of a multi-layered SIEM architecture enables trust-adaptive thresholds and continuous online model updates, resulting in more resilient and context-aware detection performance.

Our email forensics pipeline extends this lineage by fusing content-based NLP (tokenization, NER, phishing detection) with authorship-consistency profiling and AI-synthetic text probability, producing unified cognitive vectors for SIEM integration. Unlike prior isolated classifiers, we correlate these profiles with behavioral anomalies and ToM intent signals, enabling nuanced insider detection via stylistic deviations from established sender baselines.

\section{Dataset Description}
\label{sec:dataset}
\noindent\textbf{Enron Email and Spam Calibration Datasets.}
 We utilize the Enron email dataset \citep{Will2015}, which comprises approximately 500,000 messages from 150 employees of the Enron Corporation. After data pre-processing with the removal of very short emails and the parsing of the raw text of the email bodies, we were left with 422,204 emails. From these, we calculate global reference statistics such as the average sentence length and the lexical density necessary for the construction of profiles for the respective sender email addresses that have enough email communications (five minimum).

The Enron Spam Dataset \citep{Wiechmann2021} contains 33,716 labeled emails (51\% spam, 49\% ham) curated from the original Enron corpus for spam filtering benchmarks. It includes columns for \texttt{Subject}, \texttt{Message} body, \texttt{Spam/Ham} labels, and \texttt{Date}, making it suitable for NLP-based behavioral forensics in insider threat detection.
Following the pre-processing and filtering, the number of messages left was 33,218 for the insider threat detection experiments. To avoid duplicate leakages and near-duplicate contaminations, model selection adopted grouped cross-validation of normalized email text instead of a simple stratified hold-out split. Email content was represented by TF-IDF features of 10,000 terms, and we tested logistic regression, Naive Bayes, and random forests as classifiers, where logistic regression was chosen as the model for insider threat detection classification.

\noindent\textbf{Synthetic Emails for MAS Runtime Scenarios.}
In addition to the offline Enron-based calibration approach, synthetic suspicious emails were created in order to develop MAS runtime scenarios as well as simulate the communication events. The synthetic emails were created in order to resemble suspicious patterns of interest in the context of enterprises, including malicious attachments, invoice frauds, spoofing, critical security alerts, phishing requests, as well as fake surveys and information requests. The purpose of the synthetic emails was to enrich the realism of the simulations, but they did not play a part in the reporting of the grouped cross-validation results, as discussed in the following section in the context of the pretrained Enron-calibrated module.

\noindent\textbf{CERT r4.2 External Benchmark.} To complement the Mesa-based simulation, we evaluate the evidence-gated SIEM logic on the CERT r4.2 insider-threat benchmark. We use CERT as an external generalization benchmark for insider-threat detection over heterogeneous enterprise logs. The processed CERT subset contains 252 users observed over 501 days, with logon, device, file, HTTP, email, and LDAP-derived activity streams. These logs are transformed into a common user-day representation with features capturing after-hours activity, file access and copy behavior, sensitive-file indicators, suspicious HTTP activity, external email behavior, attachment use, recipient counts, and role-based peer-deviation scores. Ground-truth labels are derived from the CERT answer files at both the user-day level and actor level. CERT r4.2 is used to evaluate the generalizability of the evidence-gated SIEM logic beyond the controlled simulation environment.

\noindent\textbf{Operational Facilities Email Corpus V2.} To evaluate the domain shift between Enron-style corporate email and a different operational email domain, we created a synthetic Operational Facilities Email Corpus V2. The corpus contains 2,000 emails, including 300 phishing/suspicious emails and 1,700 benign operational emails. To reduce the risk of overly simple keyword separation, the corpus includes 300 hard benign near-miss messages that contain security-sensitive operational language, such as approved vendor remote access, approved log export, approved firmware updates, and legitimate emergency maintenance requests. The corpus also includes template identifiers and template-family labels, enabling grouped template-held-out evaluation rather than relying only on random train/test splits. All names, organizations, domains, and URLs are fictional, and the corpus is used only as a synthetic target-domain benchmark for measuring Enron-to-target-domain transfer and adaptation. The corpus is synthetic and template-generated. It is used to evaluate transfer and adaptation under operational email communication patterns, not as real-world email validation.

\section{Proposed Methodology}
\label{sec:method}
\begin{figure}[!htbp]
    \centering
    \includegraphics[width=0.75 \textwidth]{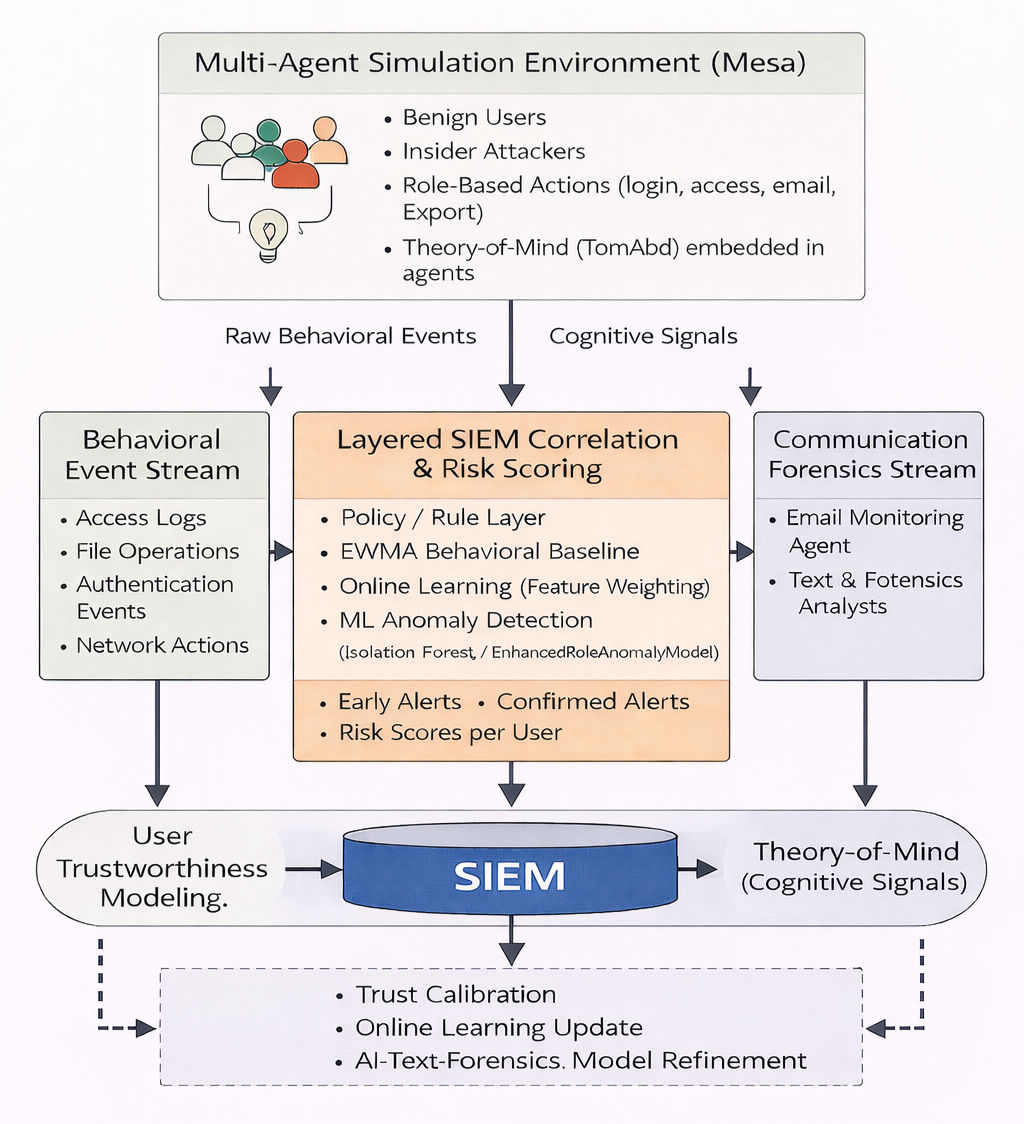}
    \caption{Overview of the proposed hybrid insider threat detection framework.
    The ML anomaly detection block names two implementations; only the role-aware
    Isolation Forest is exercised in the experiments reported here.}
    \label{fig:architecture}
\end{figure}

Our framework combines multi-agent simulation, layered correlation from SIEM, communication forensics, and cognitive trust with Theory-of-Mind (ToM) reasoning into an insider threat detection framework. As shown in Figure~\ref{fig:architecture}, an organizational simulation running Mesa generates behavioral events and intent predictions for legitimate users as well as insider attackers. Both sets of data, along with data from an intersecting communication forensics component, are fed into a centralized layered SIEM correlation and risk scoring component. The SIEM component becomes the centralized authority for both early and confirmed alarms and captures data for determining the trustworthiness of all users and model calibration feedback for improving detection performance with time. We proceed with the detailed discussion of the different components of our framework.

\subsection{Multi-Agent Simulation (MAS)}
We construct an organizational environment in Mesa \citep{terHoeven2025} where agents represent users and insider attackers. Agents have roles (staff, developer, admin) and resources they can access. A subset of agents are adversaries executing insider tactics e.g. data exfiltration, privilege escalation, colluding with external parties. The simulation runs in discrete timesteps; at each step, agents perform actions such as login, database queries, file exports, or sending emails. Crucially, each agent is embedded in a Theory-of-Mind (ToM) layer implemented via TomAbd \citep{Montes2023} which is an abductive Theory-of-Mind framework that infers latent agent beliefs, goals, and plans from observed actions. Unlike purely statistical anomaly detection, TomAbd enables plan-aware and intention-driven reasoning under partial observability, making it particularly suitable for insider threat scenarios where malicious behavior unfolds gradually and strategically. TomAbd maintains beliefs, goals, and nested plans for each agent. Observed actions feed into an abductive reasoning module that infers intentions e.g. "this user likely aims to exfiltrate sensitive configuration backups or access-control records". These inferred intentions are available as meta-features to the SIEM. In this way, the simulator produces not only raw event logs but also cognitive/contextual signals about each agent's state and predictions of next steps. The MAS can scale from dozens to thousands of agents, supporting a wide range of diverse scenarios. We use Mesa \citep{terHoeven2025} for flexibility and integration with our SIEM code.

\subsection{SIEM Layering}
\label{sec:SIEM}
All events from the MAS flow into a central SIEM component. This SIEM correlates data and issues alerts through multiple configurable layers:
\begin{itemize}
    \item \textbf{Policy layer:} This involves static policies or rules regarding violations such as the absence of access to certain servers or against the sharing of files, and this layer functions as a coarse filter, suppressing or marking certain actions, such as sending sensitive files or configuration data to external email.
     \item \textbf{Baseline anomaly layer:} This involves the use of the Exponentially Weighted Moving Average (EWMA) model to observe the usual rates for each user's activity, for example, log-in rates and data downloaded. Any deviation from the usual rates translates to a baseline score. A good example of this is when an employee who has not been actively accessing any data downloads heavily.
     \item \textbf{Trust-adaptive threshold:} One of the major contributions of the approach is that there is a user-specific, adaptively set warning threshold that is a function of the user's ``trust score.'' More concretely, the approach utilizes both an ``early warning'' threshold and a ``confirmed'' threshold. The confirmed threshold is calculated using base + slope(trust $-$ 0.5). The trust scores begin at 0.7 and evolve over time. A more trustworthy individual requires a stronger cumulative score for an alert to sound, while a less trustworthy individual needs a weaker score. This adds the notion that the record of previous alerts, true or false, impacts current alert sensitivity.
    \item \textbf{Online learning layer:} In the LSC/CE implementation, a logistic
    feature-weight update is used with handcrafted indicators such as recent email,
    after-hours login, large export, and unapproved access. In the EG-SIEM and
    EG-SIEM-Enron experiments reported here, the main adaptive components are the
    trust-updated confirmed threshold, EWMA behavioral baselines, evidence-gated
    confirmation, role-based Isolation Forest anomaly scoring, and the pretrained
    communication-forensics classifier. We therefore report the online feature-weight
    settings separately in Table~\ref{tab:hyperparameters_reproducibility} and do
    not attribute the EG-SIEM-Enron results solely to online logistic updating.
     \item \textbf{Isolation Forest anomaly detection:} Finally, for each user-role, we train an Isolation Forest on historical behavior vectors and score new activity. Outlier scores from the forest nudge borderline cases; by default we weight the ML anomaly as a ``weak advice'' that can raise a score that is near the threshold.
\end{itemize}

The SIEM maintains a sliding window of recent activity for every unique user. At every anchor point (e.g. an email or login entry point), it computes the baseline score based on weighted features, then combines this with the baseline anomaly score, and finally compares it to early and confirmed thresholds, respectively. Alerts are raised in tiers (e.g. a low-confidence early alert vs a high-confidence confirmed alert). All alerts are timestamped and labeled with user ID and context for evaluation.

In addition to basic correlation and risk scoring, we incorporate mechanisms to minimize the problem of false positives and increase the confidence of the analyst. First, regularity of behavior as a regular pattern is captured by measuring deviations from expected inter-arrival patterns and suppressing risk scores if behavior follows regular patterns. Second, peer group normalization is incorporated by comparing the export behavior of a given user to those with a similar role and scoring only those users who are significant outliers relative to their peer group. Third, for confirmed alerts, evidence accumulation is a requirement: at least two distinct types of evidence must be present before escalation occurs. Confirmed escalation involves a multi-condition gate such as Tight Exfiltration Chains, Staging Activity, Login Context, or Excess Evidence. Lastly, a compliance override is incorporated to ignore alerts indicating confirmed escalation in the presence of strong compliance factors.

The effectiveness of the system is evaluated by means of a measurable set of five insider threat scenarios that include exfiltration, stealth, takeover, staging exfiltration, and email leakage, in addition to providing metrics from the actor and alarm views. These metrics include precision, recall, F1, and Time-To-Detection (TTD).

\subsection{Email and Communication Forensics}
\begin{figure}[htbp]
    \centering
    \includegraphics[width=0.5 \textwidth]{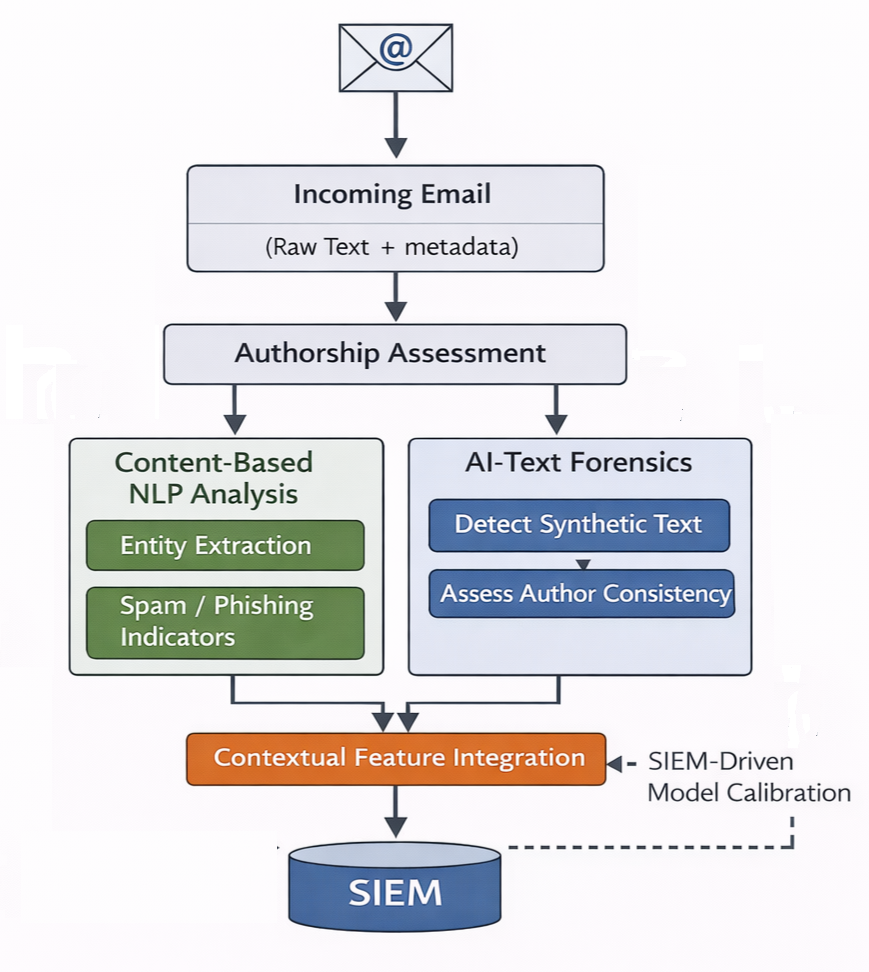}
    \caption{Email Monitoring Agent's text-mining and AI-text forensics pipeline.}
    \label{fig:textmining}
\end{figure}

Parallel to raw event analysis, the system deploys specialized forensics agents to process communication content. In the case of email traffic, an Email Monitoring Agent analyzes both the raw text and associated metadata of each incoming message. As illustrated in Figure~\ref{fig:textmining}, the email is evaluated through two complementary analysis streams operating in parallel.

The first stream carries out the content-based natural language processing (NLP) analysis by applying text-mining methodologies like tokenization, named entity identification, indicator identification, and phishing/spam categorization in accordance with the text-mining data sources utilized by the SIEM solution. The first stream identifies the semantic and functional aspects associated with the email content.

At the same time, an AI text forensics stream deals with authorship and text generation features. This assessment calculates the probability of an email being generated or supported by AI and assesses the uniformity of style with respect to past writing patterns. Instead of aiming to determine specific attribution, the forensic tool provides probabilities regarding the usage of synthetic text and style consistencies regarding the author, which serve as behavioral cues.

The results of these two streams of analyses are then fused by means of a contextual feature integration layer to produce a unified feature vector that embraces content-level markers and signals associated with authorship. This augmented metadata serves to inform the SIEM system about correlations with behavioral activity markers, usage patterns, and role-defined expectations. Thus, it becomes clear that the SIEM risk score processing would be based on email transmission activity, along with markers associated with composition and possible authorship (AI-based).

We extend the Email Monitoring Agent by introducing the pretrained forensics component, which is loaded dynamically as a serialized resource at runtime. This component is trained to operate on two differing Enron datasets: (1) the full Enron email dataset, used to establish baselines on writing styles and vocab statistics, and (2) the Enron Spam Dataset, used to train the phishing classifier component. On every outgoing email event, the component calculates: (i) phishing probability via supervised classification trained on the Enron spam/ham dataset enhanced by keyword features; (ii) urgency indicators for sensitive keywords; (iii) writing style anomaly scores according to variance in Enron baselined stats (sentence length \& lexical diversity); (iv) authorship consistency using a rolling per-actor style profile; and (v) an AI likeness indicator derived via low variance in sentence length patterns scored per sentence. These feature values are then added to the email event and aggregated over some time period. The SIEM scoring and correlation component enables a forensics flag when the recent phishing probability of the actor crosses some threshold value, using this flag as weighted evidence in alert scoring \& correlation logic. Enron datasets are used only to train/calibrate the pretrained forensics module; all emails in the MAS simulation are synthetically generated at runtime and are not drawn from Enron.

Moreover, the SIEM also provides a feedback mechanism based on alert outcomes and validation by the human analyst (for example, confirmation or flagging it as a false positive). This feedback can be used to refine the models of AI-text forensics and classification to enable improved detection rates without affecting real-time email processing.

\subsection{Theory-of-Mind and Trust Modeling}

In the MAS, agents hold a user trustworthiness score that is controlled by the SIEM, indicating the confidence of the system in the benign nature of the user. This score changes depending on the results and confirmation of alerts. If the alert is a true positive, indicating malicious or policy-breaking activity, the user's score will go down, and subsequent actions will be stringently reviewed. However, in the case of a false positive, the score increases, and subsequent actions will have less severe reviews. Concretely, on a malicious hit we apply a fixed negative delta to trust; on a false alarm, we add a positive delta. Trust is bounded between 0.10 and 0.95. In the reported EG-SIEM and
EG-SIEM-Enron experiments, trust changes only after confirmed alert outcomes:
a true positive applies a negative trust delta, while a false positive applies
a positive trust delta. Although the LSC implementation includes a decay hook,
its configured decay rate is 0.00; therefore, decay toward baseline is not
active in the reported experiments. This models an adaptive ``confidence'' in each user's benignness, which then influences the alerting threshold as described. By dynamically calibrating trust, we mimic human-AI collaboration: analysts learn to trust the system for users that turn out benign, and demand more evidence for repeat offenders.

In tandem, the ToM layer in each agent continuously infers others' intentions. For example, observing an agent performing an unusual query may lead TomAbd to infer a goal of data theft. These inferences enter the SIEM as a kind of ``cognitive clue''.
The current EG-SIEM implementation incorporates ToM intent and plan-completion
signals as weighted evidence in the SIEM score and evidence gate. However, ToM evidence alone is not sufficient to generate a confirmed alert; confirmed escalation still requires the multi-evidence confirmation conditions described in Section~\ref{sec:SIEM}. These signals help distinguish behavior that is merely unusual from behavior that is consistent with inferred malicious goals, such as exfiltration, privilege escalation, staging, or unauthorized administrative control-plane activity. Future work will extend this feature-level ToM integration toward richer coordinated-intent models and multi-actor reasoning.

\subsection{Unified Framework Formalization}
\label{sec:formalization}

Let $\mathcal{U} = \{u_1, \ldots, u_N\}$ be the set of monitored users
and $t \in \{1, \ldots, T\}$ a discrete simulation timestep. At each
step, every agent $u_i \in \mathcal{U}$ produces events consumed by the
SIEM. The framework processes these events through five sequential
stages.

\subsubsection*{Stage 1: Risk Scoring}
The SIEM aggregates evidence from all active components into a single
risk score for each user at each step:
\begin{equation}
  S_i^t = w_b B_i^t + w_a A_i^t + w_\tau \tau_i^t + w_f F_i^t + w_m M_i^t,
  \label{eq:risk}
\end{equation}
where $B_i^t$, $A_i^t$, $\tau_i^t$, $F_i^t$, and $M_i^t$ are the
behavioral deviation, anomaly, intent, forensics, and peer-normalization
scores respectively, and $w_b, w_a, w_\tau, w_f, w_m \geq 0$ are fixed
non-negative weights held constant across all reported runs. The
weights governing the behavioral, intent, and forensics terms are
listed in Appendix~\ref{app:hyperparameters}; the anomaly and
peer-normalization terms act as bounded adjustments to borderline
scores, and the complete set of configured values is released with
the implementation \citep{mas_siem_github}. Each term is defined in
the stages below.

\subsubsection*{Stage 2: Behavioral Baseline (EWMA)}
The behavioral deviation score measures normalized departure from each
user's own rolling baseline:
\begin{equation}
  B_i^t = \frac{|x_i^t - \mu_i^t|}{\sigma_i^t + \epsilon}, \qquad
  \mu_i^t = \alpha\, x_i^{t-1} + (1-\alpha)\,\mu_i^{t-1},
  \label{eq:ewma}
\end{equation}
where $x_i^t$ is the observed activity count, $\mu_i^t$ is the
EWMA mean with smoothing factor $\alpha = 0.08$, and deviations are
clipped at a z-score of $6.0$ to suppress isolated spikes.

\subsubsection*{Stage 3: Theory-of-Mind Intent Signal}
TomAbd maintains a posterior over $K$ latent goals
$\mathbf{G} = \{g_1, \ldots, g_K\}$ per agent. The intent contribution
to Eq.~\eqref{eq:risk} is:
\begin{equation}
  \tau_i^t = \max_{k}\; P(g_k \mid \mathbf{o}_i^{1:t}),
  \label{eq:tom}
\end{equation}
where $\mathbf{o}_i^{1:t}$ is the observed action sequence up to step
$t$. A contradiction factor of $0.85$ per approved contextual event
reduces $\tau_i^t$ when a benign explanation exists, preventing routine
compliant actions from inflating the intent signal.

\subsubsection*{Stage 4: Trust-Adaptive Threshold}
Each user maintains a trust score $\rho_i^t \in [0.10, 0.95]$,
initialized at $0.70$. The confirmed escalation threshold is:
\begin{equation}
  \theta_i^t = \theta_{\text{base}} + \delta_{\text{role}}(u_i)
               + 1.2\,(\rho_i^t - 0.5),
  \label{eq:threshold}
\end{equation}
where $\theta_{\text{base}} = 4.0$ and $\delta_{\text{role}}$ is a
role offset (staff $= 0$; analyst $= 0.8$; admin $= 1.2$). Trust
is updated after each confirmed outcome:
\begin{equation}
  \rho_i^t \leftarrow
  \begin{cases}
    \rho_i^{t-1} - 0.18 & \text{(confirmed true positive),} \\
    \rho_i^{t-1} + 0.05 & \text{(confirmed false positive).}
  \end{cases}
  \label{eq:trust}
\end{equation}

\subsubsection*{Stage 5: Evidence Gate}
A confirmed alert is issued for $u_i$ at step $t$ if and only if:
\begin{equation}
  S_i^t > \theta_i^t
  \;\land\;
  |\mathcal{E}_i^t| \geq 2
  \;\land\;
  \sum_{e \in \mathcal{E}_i^t} w_e \geq 2.5,
  \label{eq:gate}
\end{equation}
where $\mathcal{E}_i^t$ is the set of distinct evidence categories
accumulated in the current sliding window. The conjunction of a
score threshold, a minimum evidence-type count, and a minimum total
evidence weight ensures that no single signal can independently
trigger escalation, which is the core property that keeps confirmed
false positives near zero across all evaluated scales.

\section{Implementation and Results}
\label{sec:results}
To analyze the impact of cognitive context and precision-focused gating, four variants of the proposed framework are evaluated. Layered SIEM-Core (LSC) serves as the baseline framework that represents a traditional layered SIEM processing pipeline consisting of policy verification, trust-aware scoring, EWMA based deviation modeling, as well as deviation detection based on fixed confirmation thresholds, all of which do not make use of cognitive processing or communication evidence. Based on this starting point, Cognitive-Enriched SIEM (CE-SIEM) enhances the correlation layer with the incorporation of Theory-of-Mind (TomAbd)-based intent features and communication forensics features such as natural language processing for phishing purposes and AI text/authorship analysis. This is combined with behavioral anomalies with the same confirmation logic as the original approach. Next, Evidence-Gated SIEM (EG-SIEM) introduces precision-focused validation layers upon CE-SIEM. The variation imposes the use of evidence-gated accumulation, the use of peer-group normalization compared to role-based behavioral groups, the exclusion of routine and scheduled events, and the inclusion of Theory-of-Mind validation with awareness of contradiction. The various validation methods are expected to optimize the prevention of false positives and the number of alerts, with quick detection being ensured. Lastly, the Enron-enabled EG-SIEM (EG-SIEM-Enron) builds upon the EG-SIEM framework, dynamically loading the pretrained Enron-centric email forensics module during runtime. This component contributes normalized estimates of phishing likelihood and style-baseline deviation, all of which are combined into weighted evidence within alert scores and correlation. The proposed validation and calibration techniques are expected to reduce false positives and alert rates while preserving timely detection.

The pretrained email forensics module is calibrated offline using two Enron-derived resources: the full Enron corpus to provide reference style baselines and sender-profile statistics (422,204 emails, vocabulary size 575,895, baseline sentence length 14.60, baseline lexical richness 0.633), and the Enron spam/ham dataset to train the phishing-classification component (33,218 samples after filtering, 15,123 normalized unique groups). To reduce duplicate leakage, model selection is performed with grouped cross-validation over normalized email text rather than a simple stratified hold-out split. Using TF--IDF features with 10,000 terms, we evaluate logistic regression, naive Bayes, and random forest classifiers and select logistic regression as the runtime model based on grouped-CV performance. The selected classifier achieves mean accuracy of 0.7139, precision of 0.7462, recall of 0.9167, F1 of 0.8054, and ROC--AUC of 0.6512, with an operating threshold of 0.002. These calibration settings are listed alongside the remaining configuration in Appendix~\ref{app:hyperparameters}.

The above variants are implemented using Python Mesa to simulate the multi-agent system and incorporating customized modules for SIEM, TomAbd, and Behavioral Forensic; the implementation and experiment scripts are publicly available for reproducibility \citep{mas_siem_github}. The simulation models an organizational environment with 42 human agents: 30 benign users, 4 power users, and 8 malicious insiders across five attack scenarios, plus 4 system agents (DBMonitor, EmailMonitor, AuthMonitor, and SIEMAgent), for 46 total agents in the Mesa model. In addition to regular benign users, we model a small number of power users with privileged access consistent with administrative and operational roles, such as system operators, IT administrators, and vendor-integrator accounts. Power users exhibit a higher baseline rate of legitimate file access, database queries, configuration reads/writes, and data exports than regular users; this role supports stress-testing baseline models and peer-group normalization. In each run, a total of T = 240 simulation steps are executed, with an initial 60-step warm-up period for developing baselines and building anomaly models based on roles. Post-warm-up, the testing phase begins with insider activities enabled. In each run, the SIEM system ingests monitored events and produces two levels of alerts: Early Alerts for Triage and Confirmed Alerts for Escalation. Detailed hyperparameters and reproducibility settings are provided in Appendix~\ref{app:hyperparameters}.

Ground-truth labels are generated directly by the simulator based on the assigned role and actions of actors. Evaluated metrics are: (i) actor precision, recall, and F1, considering an actor to be detected whenever they generate at least one confirmed alarm in the testing phase, (ii) confirmed-alert precision, and (iii) time to detection, calculated from the first malicious activity to the first confirmed alarm.

\subsection{Quantitative Results and Trade-offs}
\label{sec:quantitative}

Table~\ref{tab:quant_results} summarizes the averaged performance over 10 runs for the four configurations. For the LSC baseline, we report the operating point with confirmed threshold $\theta=4$, which yields a confirmed-alert volume of 107.3/run; we use this as the matched-baseline operating point for the four-variant comparison. LSC achieves actor-level F1=0.567 (P=0.405, R=0.950) with confirmed-alert precision 0.680. Although recall is high, the baseline generates substantial false positives (33.7 confirmed FP/run), reflecting limited discrimination when cognitive and communication-derived evidence is not available. Adding Theory-of-Mind and communication forensics in CE-SIEM increases actor-level performance to F1=0.774 (P=0.633, R=1.000), representing a large gain over LSC. This improvement is achieved by detecting all insider actors (R=1.0) but comes with a higher operational burden: CE-SIEM produces 152.0 confirmed alerts per run and confirmed-alert precision of 0.677, indicating a stronger but noisier detection posture.

EG-SIEM substantially improves reliability and reduces alert fatigue. Actor-level F1 rises to 0.898 (P=0.973, R=0.838) while confirmed-alert precision reaches 0.994 with only 36.2 confirmed alerts per run and 0.2 confirmed FP/run. Compared to CE-SIEM, EG-SIEM cuts confirmed alert volume by approximately 76\% while improving confirmed-alert precision from 0.677 to 0.994. The trade-off is a reduction in recall and a substantial increase in average TTD (52.98 vs. 15.68 steps for CE-SIEM), consistent with the evidence gate requiring multiple corroborating signals before escalation. Worst-case TTD also rises (max 145.20 vs. 61.00 steps), reflecting the deliberate design choice to delay confirmation until high-confidence evidence accumulates. Finally, EG-SIEM-Enron enhances EG-SIEM by adding a trained email forensics module based on the Enron corpus. This provides a phishing likelihood measure as well as a measure of deviation from the email style baselines to be used as additional weighted evidence. The parameters keep a conservative alert policy (0.0 confirmed FP/run; confirmed-alert prec. 1.000) but offer a small increment to actor-level F1 to 0.944 (P=1.000, R=0.900). Average TTD increases further to 82.81 steps (max 256.00), reflecting the additional evidence accumulation required by the email-forensics component before a confirmed alert is issued. This confirms the core precision--latency trade-off of the evidence-gated design: adding a richer forensics signal eliminates all confirmed false positives at the cost of a longer confirmation window.

To test whether the ordering in Table~\ref{tab:quant_results} is statistically
supported rather than an artifact of seed variance, we compared LSC, CE-SIEM, and
EG-SIEM using paired Wilcoxon signed-rank tests on run-level actor F1 across the ten
shared seeds, applying Holm--Bonferroni correction across the three comparisons. All
three differences remain significant after correction: LSC versus CE-SIEM (mean
difference $+0.207$, $W = 0.0$, $p = 0.0020$, Holm-adjusted $p = 0.0059$), LSC versus
EG-SIEM ($+0.331$, $W = 0.0$, $p = 0.0020$, Holm-adjusted $p = 0.0059$), and CE-SIEM
versus EG-SIEM ($+0.123$, $W = 3.0$, $p = 0.0098$, Holm-adjusted $p = 0.0098$). The
non-parametric test is used because actor F1 over ten runs is bounded and not assumed
normal. EG-SIEM-Enron is excluded from the paired analysis because its
forensics-primary runs are not archived with the per-seed identifiers required to
pair them against the other three variants, and a paired signed-rank test requires
seed-level pairing; the aggregate statistics reported for it in
Table~\ref{tab:quant_results} are unaffected. Its comparison with EG-SIEM should
therefore be read as descriptive rather than as a tested difference.

\begin{table}[!htbp]
\centering
\caption{Actor-level detection and alert statistics across SIEM variants.
         Means and standard deviations are computed over ten matched runs
         (seeds 42--51, 240 steps/run, 60-step warm-up). EG-SIEM-Enron uses
         the forensics-primary full-mode configuration. 95\% confidence
         intervals on actor F1 are derived from the $t$-distribution ($n=10$).}
\label{tab:quant_results}
\small
\setlength{\tabcolsep}{3.5pt}
\begin{tabular}{lcccc}
\hline
\textbf{Metric} & \textbf{LSC} & \textbf{CE-SIEM}
                & \textbf{EG-SIEM} & \textbf{EG-SIEM-Enron} \\
\hline
Actor Precision
  & $0.405 \pm 0.044$ & $0.633 \pm 0.043$
  & $0.973 \pm 0.057$ & $1.000 \pm 0.000$ \\
Actor Recall
  & $0.950 \pm 0.065$ & $1.000 \pm 0.000$
  & $0.838 \pm 0.084$ & $0.900 \pm 0.118$ \\
Actor F1
  & $0.567 \pm 0.050$ & $0.774 \pm 0.034$
  & $0.898 \pm 0.063$ & $0.944 \pm 0.068$ \\
\quad 95\% CI on F1
  & {[}0.531,\,0.602{]} & {[}0.750,\,0.798{]}
  & {[}0.853,\,0.943{]} & {[}0.895,\,0.992{]} \\
\hline
TTD avg (steps)
  & $20.46 \pm 14.79$ & $15.68 \pm 5.49$
  & $52.98 \pm 19.38$ & $82.81 \pm 32.78$ \\
TTD max (steps)
  & $69.80 \pm 51.68$ & $61.00 \pm 27.81$
  & $145.20 \pm 56.54$ & $256.00 \pm 115.39$ \\
\hline
Confirmed alerts (avg/run)
  & $107.3 \pm 15.2$ & $152.0 \pm 8.7$
  & $36.2 \pm 3.5$ & $65.8 \pm 6.8$ \\
Confirmed-alert precision
  & $0.680 \pm 0.055$ & $0.677 \pm 0.033$
  & $0.994 \pm 0.012$ & $1.000 \pm 0.000$ \\
Confirmed FP (avg/run)
  & $33.7 \pm 2.75$ & $49.3 \pm 7.54$
  & $0.2 \pm 0.42$ & $0.0 \pm 0.00$ \\
\hline
ToM-assisted detections (avg)
  & -- & 111.2 & 31.9 & 0.0\\
\hline
\end{tabular}
\begin{flushleft}
\footnotesize
\textit{Note:} The ToM-assisted detection row is a diagnostic counter, not a measure
of the ToM contribution to detection, and the counts are not comparable across
variants; see the discussion below. LSC contains no ToM module, so the entry is not
applicable.
\end{flushleft}
\end{table}

\begin{figure}[!htbp]
\centering
\subfloat[Actor F1 distributions.\label{fig:f1_boxplot}]{
  \includegraphics[width=0.48\textwidth]{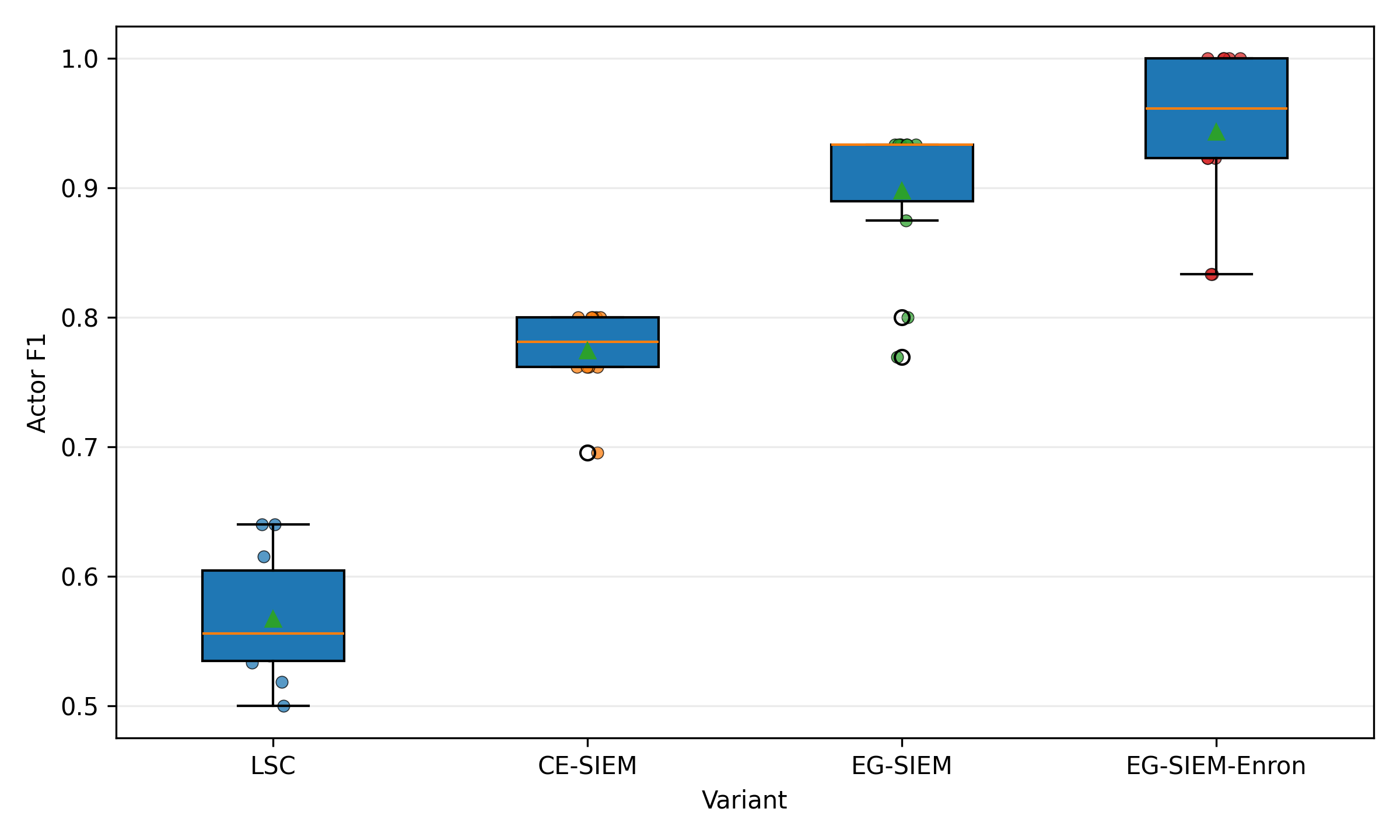}
}
\hfill
\subfloat[Average time-to-detection distributions.\label{fig:ttd_boxplot}]{
  \includegraphics[width=0.48\textwidth]{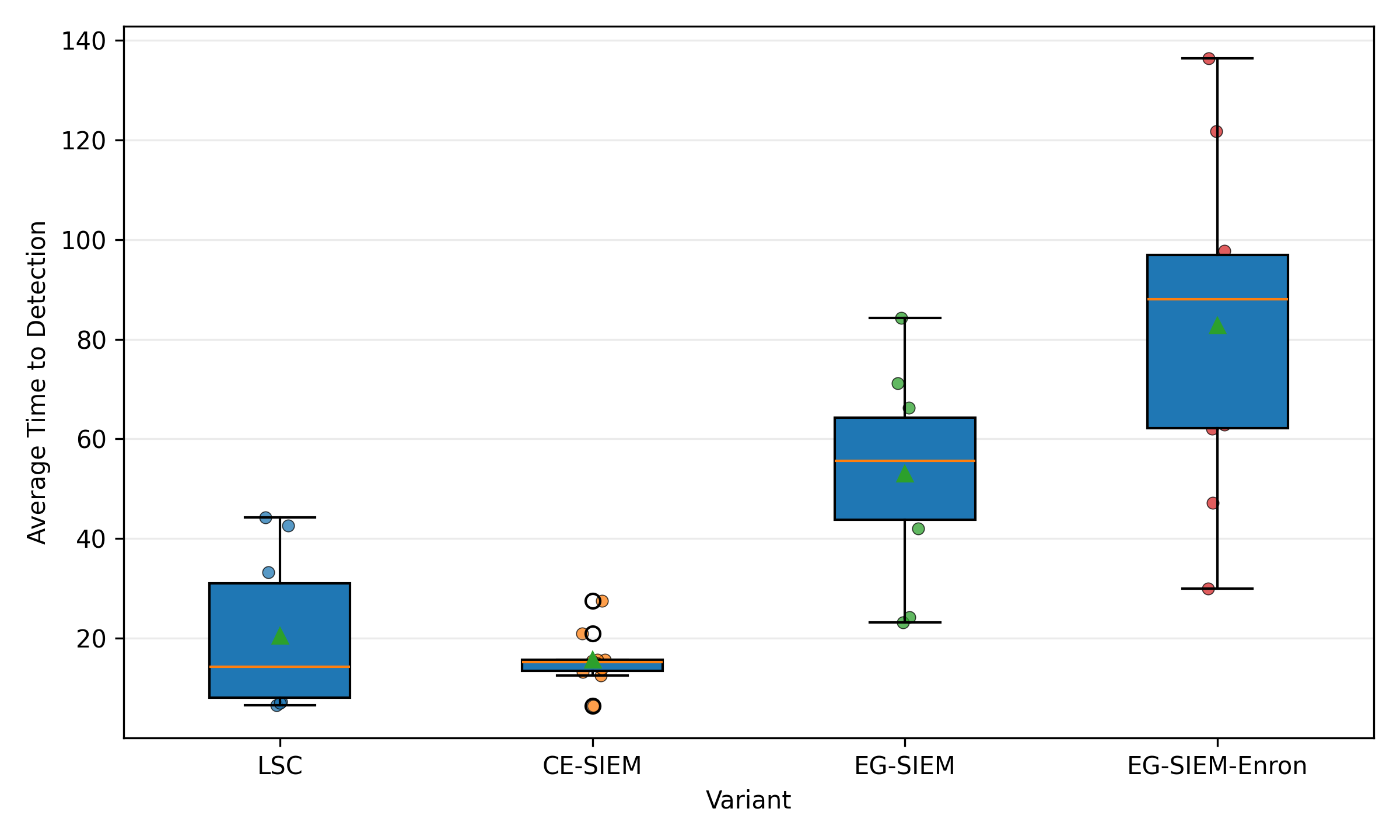}
}
\caption{Run-level distributions across ten matched seeds (42--51) for LSC, CE-SIEM, EG-SIEM, and EG-SIEM-Enron. Individual run values are shown as scatter points.}
\label{fig:matched_boxplots}
\end{figure}

Figure~\ref{fig:matched_boxplots} shows the run-level actor F1 and average time-to-detection distributions across the ten matched seeds.

\begin{figure}[!htbp]
\centering
\includegraphics[width=\columnwidth]{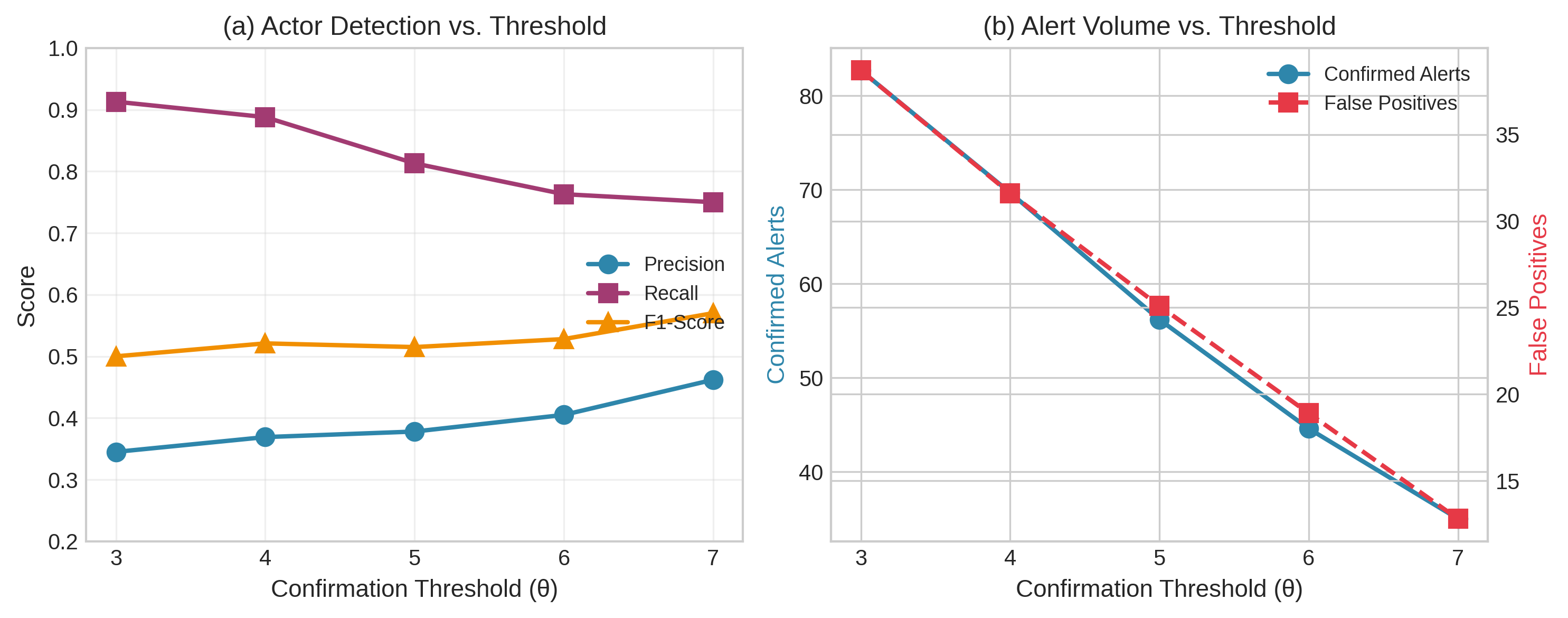}
\caption{LSC sensitivity to confirmed threshold $\theta$ (10-run averages).}
\label{fig:lsc_sweep}
\end{figure}

Figure~\ref{fig:lsc_sweep} presents the LSC baseline performance across different confirmation thresholds ($\theta$ = 3 to 7). The analysis reveals the inherent precision-recall trade-off in traditional SIEM approaches. Increasing the confirmation threshold improves precision (from 0.345 at $\theta=3$ to 0.462 at $\theta=7$) but degrades recall (from 0.913 to 0.750) and increases detection latency (average TTD rises from 13.09 to 39.13 steps). The sweep values represent averages over a separate set of runs and differ slightly from the 10-run matched comparison in Table~\ref{tab:quant_results} due to stochastic variance. This trade-off demonstrates the fundamental limitation of threshold-based approaches: without cognitive context or evidence gating, the system cannot simultaneously achieve high precision and rapid detection.

The TTD values should be interpreted in relation to the simulation timestep rather than as fixed wall-clock seconds. In an operational deployment, the real-time implication depends on the SIEM ingestion interval, the event polling interval, and the SOC escalation policy. The evidence-gated design intentionally delays confirmed escalation until multiple evidence streams accumulate, reducing false positives but creating a trade-off between speed and confidence. For critical control actions, the framework should therefore be deployed with fast policy-layer blocking or step-up authentication for high-risk actions, while evidence-gated confirmation supports analyst triage and investigation. Thus, confirmed TTD reflects the time to high-confidence escalation, not necessarily the first opportunity for preventive control.

A run-level view of EG-SIEM adds detail to the aggregate figures in
Table~\ref{tab:quant_results}: the precision gain is consistent rather than driven by
a few favorable seeds, since eight of the ten runs achieved perfect actor precision
with zero confirmed false positives, and only runs 7 and 8 produced a single false
positive each, which is what yields the 0.2 average. Across the same runs, the
TomAbd module registered an average of 111.2 assisted detections per run in CE-SIEM,
31.9 in EG-SIEM, and 0.0 in EG-SIEM-Enron.

These counts should be read as a diagnostic of how often the ToM signal was recorded
as contributing, not as a measure of the ToM contribution to detection, and they are
not comparable across variants. Two reasons apply. First, the counter reports 0.0 for
every forensics-primary run, including the model-only and full modes in
Table~\ref{tab:enron_mode_ablation} that do produce confirmed alerts; the zero entry
for EG-SIEM-Enron therefore reflects how the counter is exercised under the
forensics-primary preset and is not a measurement that ToM ceases to contribute.
Second, ToM enters as one weighted category among the eight listed in
Appendix~\ref{app:hyperparameters}, and the evidence gate can be satisfied by pairing
communication-forensics evidence with a behavioral category without the ToM intent
term, so a low count is compatible with ToM having shaped the underlying risk score.
The corresponding scoping limitation is stated in Section~\ref{sec:limit}.

\subsection{Detection Quality at Larger Simulation Scales}
\label{sec:scalability}
To verify that the actor-level results in Table~\ref{tab:quant_results} are not dependent on the 42-agent setting, we re-evaluated the unchanged EG-SIEM evidence-gated configuration over larger populations and longer horizons, without retuning any thresholds or weights and repeating each configuration three times. Table~\ref{tab:mesa_scalability} reports detection metrics under populations of 100, 250, 500, and 1000 human agents at 240 and 480 simulation steps. First, actor-level F1 remains within a narrow band of 0.875--0.933 across all scales, a degradation of approximately five percentage points over a 24-fold increase in population. Second, the runtime profile scales acceptably: peak memory grows approximately linearly with the number of agents (6.6\,MB to 158.5\,MB), runtime grows sub-linearly with agent count for a fixed step count (43.7\,s to 158.0\,s), and event throughput rises from roughly 99 to 1{,}234 events/second. Third, confirmed false positives remain at most one per run even at 500--1000 agents, indicating that the evidence-gated logic does not collapse under heavier workloads. The 1000-agent setting completed one of three runs within the 16\,GB memory guard; we report this transparently rather than dropping it. Together, these results support the claim that the detection quality reported in Table~\ref{tab:quant_results} reflects properties of the evidence-gated framework rather than small-scale artifacts.

\begin{table}[!htbp]
\centering
\caption{Mesa simulation scalability under the unchanged EG-SIEM evidence-gated
         configuration (no threshold or weight retuning).}
\label{tab:mesa_scalability}
\small
\begin{tabular}{rrrccc}
\toprule
\textbf{Agents} & \textbf{Steps} & \textbf{Runs} &
\textbf{Actor F1} & \textbf{Conf.\ alerts} & \textbf{Conf.\ FP} \\
\midrule
   42 & 240 & 3 & 0.933 &      38.0 & 0.0 \\
  100 & 240 & 3 & 0.875 &      78.0 & 0.0 \\
  100 & 480 & 3 & 0.889 &     172.3 & 0.0 \\
  250 & 480 & 3 & 0.878 &     370.3 & 0.3 \\
  500 & 480 & 3 & 0.882 &     695.7 & 1.0 \\
 1000 & 480 & 1 & 0.886 & 1{,}104.0 & 1.0 \\
\bottomrule
\end{tabular}
\begin{flushleft}
\footnotesize
\textit{Note:} Values are means over completed runs. The scalability sweep was
executed by a separate runner over three seeds per setting, so the 42-agent row is
not expected to reproduce the EG-SIEM column of Table~\ref{tab:quant_results}, which
is averaged over ten matched seeds by the main comparison runner. The 1000-agent
setting completed one of three attempted runs within the 16\,GB memory guard. Runtime,
peak memory, and event-throughput measurements for each configuration are
summarized in the text and reported in full in the public repository
\citep{mas_siem_github}.
\end{flushleft}
\end{table}

\subsection{Email Forensics Ablation}

This ablation study is a component-level analysis of the EG-SIEM-Enron email forensics module in a forensics-focused setting; it is not meant to replace the main cross-variant comparison in Table~\ref{tab:quant_results}. Table~\ref{tab:enron_mode_ablation}'s \textit{Full} mode corresponds to the same EG-SIEM-Enron forensics-primary configuration reported in Table~\ref{tab:quant_results}; therefore, the EG-SIEM-Enron values in Table~\ref{tab:quant_results} and the \textit{Full} mode values in Table~\ref{tab:enron_mode_ablation} are intentionally aligned.

EG-SIEM-Enron extends the evidence-gated framework with the email forensics module calibrated on the Enron dataset with grouped cross-validation on TF-IDF features. As shown in Table~\ref{tab:enron_mode_ablation}, the hybrid full mode achieves the
highest performance of the four operating modes, with actor precision = 1.000, actor
recall = 0.900, and actor F1 = 0.944. Model-only operation reaches actor recall =
0.600 and actor F1 = 0.736, and the disabled mode produces no detections. The
keyword-only column reports actor recall = 0.414 and actor F1 = 0.553, but that run
excluded the takeover insider scenario from the simulated population, so its recall is
computed over a different insider composition and is not directly comparable with the
other three modes. The comparison that carries the argument is therefore between
model-only and full: adding the keyword layer to the trained classifier raises actor
recall from 0.600 to 0.900 and actor F1 from 0.736 to 0.944, so the hybrid
configuration outperforms the trained model operating alone. The keyword-only figures
are reported for completeness under the population caveat noted in the table.

\begin{table}[t]
\centering
\caption{Ablation of EG-SIEM-Enron email-forensics modes under the forensics-primary setting (10 runs, 240 steps/run, 60-step warmup).}
\label{tab:enron_mode_ablation}

\scriptsize
\setlength{\tabcolsep}{2.5pt}
\renewcommand{\arraystretch}{1.12}

\begin{tabularx}{\columnwidth}{
>{\raggedright\arraybackslash}X
c c c c}
\toprule
\textbf{Metric} &
\textbf{Disabled} &
\shortstack{\textbf{Keyword}\\\textbf{only}$^{\ddagger}$} &
\shortstack{\textbf{Model}\\\textbf{only}} &
\textbf{Full} \\
\midrule

Actor Precision
& 0.000 & 0.900 & 1.000 & 1.000 \\

Actor Recall
& 0.000 & 0.414 & 0.600 & 0.900 \\

Actor F1
& 0.000 & 0.553 & 0.736 & 0.944 \\

\midrule

TTD avg. (steps)
& 0.00 & 67.80 & 114.84 & 82.81 \\

TTD max. (steps)
& 0.00 & 140.00 & 274.00 & 256.00 \\

\midrule

Confirmed alerts (avg./run)
& 0.0 & 19.0 & 32.2 & 65.8 \\

Confirmed-alert precision
& 0.000 & 0.900 & 1.000 & 1.000 \\

Confirmed FP (avg./run)
& 0.0 & 0.0 & 0.0 & 0.0 \\

\midrule

ToM-assisted detections (avg.)
& 0.0 & 0.0 & 0.0 & 0.0 \\

\bottomrule
\end{tabularx}

\begin{flushleft}
\footnotesize
\textit{Note:} $^{\ddagger}$The keyword-only run was executed with the takeover
insider scenario excluded from the simulated population. Its actor recall is
therefore computed over a different insider composition from the other three modes,
and the recall gap between keyword-only and the remaining modes should not be
attributed to the forensics mode alone. The disabled, model-only, and full modes
share the standard eight-insider composition described in Section~\ref{sec:results}.
\end{flushleft}

\end{table}

\subsection{Domain-Shift Evaluation on Operational Facilities Email Corpus V2}
To directly address the concern that an Enron-calibrated email-forensics module may not transfer to a different operational email domain, we evaluated the Enron-trained TF--IDF logistic-regression classifier on the Operational Facilities Email Corpus V2. This experiment is separate from the Mesa runtime experiment. Its purpose is to measure whether Enron-derived phishing probabilities and thresholds transfer to target-domain operational emails, and whether target-domain calibration or fine-tuning is needed. Table~\ref{tab:municipal_v2_domain_shift} summarizes the five evaluation
settings.

The zero-shot result confirms a substantial domain-shift problem. Using the Enron F1-optimized threshold of 0.98, the classifier produced no positive predictions on the V2 corpus, yielding precision = 0.000, recall = 0.000, and F1 = 0.000. The default threshold of 0.50 also produced no positive predictions. A permissive Mesa runtime threshold of 0.002 recovered recall but over-flagged the entire target corpus, with predicted-positive rate = 1.000 and F1 = 0.261. Target-domain threshold calibration also recovered high recall but still over-flagged most messages, with predicted-positive rate = 0.944 and F1 = 0.261. These results indicate that threshold calibration alone is not sufficient for a deployable standalone email classifier.

Fine-tuning on target-domain examples substantially improved performance. Under random split evaluation, target-domain fine-tuning achieved F1 = 1.000. More importantly, grouped template-held-out evaluation remained strong, with F1 = 0.974 when grouped by template family and F1 = 0.993 when grouped by template ID. Since the corpus is synthetic and template-generated, these results should be interpreted as evidence that target-domain adaptation is feasible, not as real-world deployment validation. Overall, the domain-shift experiment confirms that Enron can serve as an initial calibration source, but operational deployment requires target-domain threshold calibration and fine-tuning.

\begin{table}[!htbp]
\caption{Operational Facilities Email Corpus V2 domain-shift and adaptation results.}
\label{tab:municipal_v2_domain_shift}
\centering
\small
\begin{tabular}{lcccc}
\toprule
\textbf{Evaluation Setting} &
\textbf{Precision} &
\textbf{Recall} &
\textbf{F1} &
\textbf{PR-AUC} \\
\midrule
Zero-shot Enron threshold & 0.000 & 0.000 & 0.000 & 0.128 \\
Target-calibrated threshold & 0.152 & 0.953 & 0.261 & 0.128 \\
Fine-tuned random split & 1.000 & 1.000 & 1.000 & 1.000 \\
Fine-tuned grouped by template family & 0.949 & 1.000 & 0.974 & 1.000 \\
Fine-tuned grouped by template ID & 0.986 & 1.000 & 0.993 & 1.000 \\
\bottomrule
\end{tabular}
\begin{flushleft}
\footnotesize
\textit{Note:} The Operational Facilities Email Corpus V2 is synthetic. Grouped template-held-out results are emphasized because they reduce the risk that random splits inflate performance through repeated template structure. The zero-shot row uses the Enron F1-optimized threshold of 0.98; the random-split row is reported for completeness and is likely optimistic.
\end{flushleft}
\end{table}

The hard-negative analysis further supports the need for fine-tuning rather than simple threshold adjustment. Target threshold calibration alone produced a high false-positive rate on hard benign near-miss messages, while grouped fine-tuning reduced false positives on these legitimate but security-sensitive operational emails. This finding is important for operational settings because many benign IT and operations workflows legitimately contain terms such as access, override, export, firmware, logs, VPN, or emergency. A deployable email-forensics module therefore requires target-domain adaptation that can distinguish malicious requests from approved operational workflows.

\subsection{External CERT r4.2 Benchmark Evaluation}
\label{sec:cert}

To test whether the evidence-gated correlation strategy generalizes beyond the controlled Mesa simulation, we conducted an external benchmark evaluation using CERT r4.2. This experiment is reported separately from the Mesa simulation because CERT is a real enterprise insider-threat dataset, which lets us evaluate external generalization of the SIEM correlation logic under a larger, heterogeneous, and highly imbalanced log setting.

The processed CERT subset contains 76,622 user-day records, of which 966 are malicious user-days. Thus, malicious activity represents only 1.26\% of the user-day records. Because of this severe class imbalance, user-day F1 is expected to be low. We therefore report both user-day metrics and actor-level metrics. Actor-level detection is operationally important because an analyst's practical goal is often to identify the insider account for investigation, rather than to classify every individual user-day correctly.

Table~\ref{tab:cert_external_benchmark_results} reports the full comparison. CERT-EG-SIEM full achieved actor-level precision = 0.576, recall = 0.814, and F1 = 0.675. This outperformed CERT-LSC (actor F1 = 0.491), One-Class SVM (actor F1 = 0.450), and Isolation Forest (actor F1 = 0.609). CERT-EG-SIEM full also reduced false positives from 79.72 FP/day under CERT-LSC to 6.32 FP/day. These results indicate that the evidence-gated strategy remains useful beyond the controlled Mesa environment, although the lower absolute F1 compared with the Mesa results confirms that real benchmark settings are substantially more difficult.

\begin{table}[!htbp]
\caption{External CERT r4.2 benchmark results.}
\label{tab:cert_external_benchmark_results}
\centering
\scriptsize
\setlength{\tabcolsep}{4pt}
\begin{tabular}{lccccccc}
\toprule
\textbf{Method} &
\shortstack{\textbf{User-day}\\\textbf{F1}} &
\textbf{ROC-AUC} &
\textbf{PR-AUC} &
\textbf{FP/day} &
\shortstack{\textbf{Actor}\\\textbf{Precision}} &
\shortstack{\textbf{Actor}\\\textbf{Recall}} &
\shortstack{\textbf{Actor}\\\textbf{F1}} \\
\midrule
Isolation Forest
& 0.042 & 0.774 & 0.028 & 15.08 & 0.460 & 0.900 & 0.609 \\

One-Class SVM
& 0.021 & 0.419 & 0.011 & 7.57 & 0.329 & 0.714 & 0.450 \\

Embedding-OC-SVM$^\dagger$ (all-MiniLM-L6-v2)
& 0.000 & 0.485 & 0.003 & 0.42 & 0.034 & 0.028 & 0.031 \\

CERT-LSC
& 0.034 & 0.684 & 0.027 & 79.72 & 0.326 & 1.000 & 0.491 \\

CERT-EG-SIEM without email evidence
& 0.075 & 0.768 & 0.041 & 6.30 & 0.570 & 0.814 & 0.671 \\

CERT-EG-SIEM email only
& 0.144 & 0.568 & 0.091 & 0.70 & 0.429 & 0.600 & 0.500 \\

CERT-EG-SIEM without behavioral-chain evidence
& 0.088 & 0.784 & 0.061 & 4.41 & 0.527 & 0.829 & 0.644 \\

CERT-EG-SIEM full
& 0.088 & 0.776 & 0.046 & 6.32 & 0.576 & 0.814 & 0.675 \\

\bottomrule
\end{tabular}
\begin{flushleft}
\footnotesize
\textit{Note:} CERT-LSC denotes the rule/threshold-based CERT baseline. CERT-EG-SIEM full denotes the evidence-gated CERT configuration using behavioral, email, and sequence-based evidence. The term behavioral-chain evidence is used for CERT because CERT contains observable logs only, not agent-internal Theory-of-Mind states. $^\dagger$Embedding-OC-SVM was trained on a 20{,}000-row stratified sample due to RBF SVM scalability constraints; results are not directly comparable to full-dataset rows.
\end{flushleft}
\end{table}

The CERT ablation results also clarify the role of email evidence. The full CERT-EG-SIEM model achieved actor F1 = 0.675, while the configuration without email evidence achieved actor F1 = 0.671. This shows that, in the CERT benchmark, the primary benefit comes from behavioral and sequence-based evidence rather than email alone. The email-only configuration produced the highest user-day precision among the CERT configurations (0.225) but weaker actor-level coverage (actor F1 = 0.500). Therefore, email should be understood as one complementary evidence stream, not as the sole basis for insider detection.

We further added a representation-learning
baseline in which each CERT user-day feature vector was converted to a
short behavioral text description and encoded with \texttt{all-MiniLM-L6-v2},
a compact sentence-transformer model in the Sentence-BERT family
\citep{Reimers2019SBERT}; a One-Class SVM was then trained on benign
embeddings and the top 1\% anomaly scores were flagged.
This baseline yields near-zero false positives (0.42~FP/day) but
effectively zero detection (actor F1 = 0.031, user-day F1 = 0.000),
confirming that generic sentence-encoder representations do not capture
the anomaly geometry present in tabular behavioral features, and that
CERT-EG-SIEM full retains a substantial advantage in actor-level
detection ($\sim\!22\times$ higher actor F1). This result concerns generic frozen
sentence embeddings only. It does not speak to task-specific fine-tuned LLM detectors
such as those of \citet{Song2025Confront} and \citet{Song2024AuditLLM}, which are
addressed as a scoping limitation in Section~\ref{sec:limit}.

\section{Limitations}
\label{sec:limit}
Even though the proposed framework exhibits impressive performance over all the considered SIEM approaches, there exist some limitations that must be acknowledged. Firstly, as indicated earlier, the evaluation process relies on a simulated environment rather than actual enterprise SOC network logs. The principal evaluation environment uses 42 human agents and 240 steps, which is intentionally small for controlled comparison across the four SIEM variants. Section~\ref{sec:scalability} (Table~\ref{tab:mesa_scalability}) reports an additional scalability sweep up to 1000 agents and 480 simulation steps under the unchanged EG-SIEM evidence-gated configuration, in which actor-level F1 remains within 0.875--0.933 and confirmed false positives stay at most one per run. The framework is not yet validated for the 24/7 continuous logging regime of a
real enterprise SOC; that regime requires distributed deployment and is left as future work. Even with this evidence, the simulation does not capture all operational complexities of a deployed enterprise SOC, including incomplete logging, intermittent device failures, operator interruptions, slow workflow drift, and adaptive adversaries aware of the detection logic, so the reported numbers should still be interpreted as feasibility evidence rather than deployment-grade performance.

The role of the Theory-of-Mind layer is reported here at the level of the four-variant comparison rather than as an isolated causal ablation. ToM-derived intent and plan-completion signals enter the SIEM as one weighted evidence source among several, and ToM evidence alone cannot produce a confirmed alert because confirmed escalation still requires the minimum evidence-count and evidence-weight conditions described in Section~\ref{sec:SIEM}. The current implementation does not expose a toggle that isolates the ToM contribution while holding every other CE-SIEM component fixed, and LSC contains no native ToM module against which a clean LSC+ToM condition could be constructed. We therefore do not claim a measured standalone effect for ToM, and a matched ToM ablation under identical seeds, populations, and evidence weights remains future work.

Another limitation concerns the email-forensics component and the domain shift between Enron-style corporate email and a different operational email domain. The Enron corpus and Enron spam/ham data are useful as initial calibration sources, but the Operational Facilities Email Corpus V2 experiment shows that the Enron-optimized decision threshold does not transfer directly to target-domain operational messages. Zero-shot transfer under the Enron threshold produced F1 = 0.000 on the synthetic V2 corpus. Target threshold calibration recovered recall but over-flagged many benign operational messages. Fine-tuning on target-domain examples substantially improved performance, including under grouped template-held-out evaluation. These results indicate that Enron-derived email forensics should be treated as a complementary and adaptable evidence stream rather than as a standalone email detector. Because the V2 corpus is synthetic and template-generated, future work should validate the email-forensics module on real or anonymized enterprise SOC email, ticketing, and operational communication records.

The comparative evaluation in Section~\ref{sec:cert} covers tabular unsupervised
detectors and a frozen sentence-encoder representation, but it does not include a
fine-tuned or prompted LLM detector of the kind reported by \citet{Song2024AuditLLM},
\citet{Li2025RedChronos}, and \citet{Song2025Confront}. Those systems require
substantially different training budgets and, in several cases, different CERT releases
and evaluation units, so a like-for-like comparison would require a matched
re-implementation on CERT r4.2 under the user-day and actor-level protocol used here.
We therefore position the embedding baseline as a representation-learning reference
point rather than as a state-of-the-art LLM comparison, and we treat the matched LLM
benchmark as future work.

Compliance-context handling is evaluated in this work as rule-based contradiction handling and risk-score reduction, not as a complete
deployment-grade compliance override. The present experiments do not include an adversarial compliance-exploiter scenario, independent approval workflow, immutable audit-log enforcement, or time-bounded change-window validation. These safeguards are necessary before operational deployment and remain part of future work.

The CERT benchmark strengthens the evaluation by testing the evidence-gated SIEM logic on a larger external insider-threat dataset. Because CERT is a separate enterprise dataset rather than the controlled simulation environment, these results should be interpreted as external generalization evidence rather than direct deployment validation. The lower user-day F1 observed in CERT reflects the much more severe class imbalance of the benchmark, where only 966 of 76,622 user-day records are labeled malicious. For this reason, the CERT evaluation reports both user-day and actor-level metrics. The actor-level results are more operationally meaningful for insider-threat investigation, while user-day metrics provide a stricter view of temporal localization. Future work should validate the framework using real or anonymized enterprise SOC logs and operational communication records.

Finally, the conclusions drawn in this work have been evaluated at different levels. Table~\ref{tab:quant_results} provides a matched comparison of the four major system variants, whereas the email forensics ablation study focuses on the operating modes of the Enron-enabled component in a separate email forensics environment. While the two studies are informative, they do not address the same questions and should not be read as a single continuous comparison. Furthermore, the work reported here does not yet include any real analyst-in-the-loop validation, long-horizon adaptive attackers, large-scale coordinated collusion, or open-world robustness.

\section{Conclusion}
\label{sec:concl}
This paper presented a hybrid insider threat detection framework integrating multi-agent simulation, layered SIEM correlation, trust-adaptive thresholds, Theory-of-Mind intent signals, and communication forensics for enterprise and organizational environments. The four-variant comparison across ten matched runs demonstrates a clear precision--latency trade-off. LSC achieves actor-level F1 = 0.567 with 33.7 confirmed false positives per run. Adding cognitive and communication evidence in CE-SIEM raises F1 to 0.774 and achieves perfect recall, but increases alert volume. EG-SIEM substantially reduces false positives to 0.2 per run and raises confirmed-alert precision to 0.994 (actor F1 = 0.898), at the cost of longer high-confidence confirmation time (TTD avg 52.98 steps). EG-SIEM-Enron further eliminates false positives entirely (actor F1 = 0.944, 0.0 confirmed FP/run), with TTD rising to 82.81 steps as the email-forensics component requires additional evidence accumulation before escalation. Paired Wilcoxon signed-rank tests on run-level actor F1 confirm that all three pairwise differences among LSC, CE-SIEM, and EG-SIEM remain significant after Holm--Bonferroni correction. These results indicate that evidence-gated confirmation with communication forensics is the most effective configuration for precision-critical operational settings where false-alarm cost is high.
Domain-shift evaluation on the Operational Facilities Email Corpus V2 confirms that Enron-calibrated email forensics cannot be applied zero-shot to a different operational email domain (F1 = 0.000 under the Enron threshold), but target-domain fine-tuning recovers strong performance (F1 = 0.974 under grouped template-family evaluation). External evaluation on CERT r4.2 shows that the evidence-gated strategy generalises beyond the simulation: CERT-EG-SIEM full achieves actor-level F1 = 0.675 and reduces false positives from 79.72 to 6.32 FP/day, outperforming tabular One-Class SVM, Isolation Forest, and a representation-learning embedding baseline. Mesa scalability tests from 42 to 1000 agents further confirm that actor-level F1 remains within 0.875--0.933 with at most one confirmed false positive per run, indicating that the detection quality is not a small-scale artifact.
Future work will pursue richer multi-actor Theory-of-Mind reasoning together with a matched causal ablation of the ToM layer, a direct comparison of the current role-aware Isolation Forest anomaly layer against richer temporal anomaly models, validation against real or anonymised enterprise SOC logs and operational telemetry, longer-horizon adaptive and collusive insider scenarios, and federated or privacy-preserving deployment architectures suitable for operational SOC environments.


\section*{Disclosure statement}
No potential conflict of interest was reported by the author(s).

\section*{Data availability statement}
The simulation code, SIEM implementation, experiment scripts, and configuration
files that support the findings of this study are openly available in the
Insider-Threat-Detection-MAS-SIEM repository at
\url{https://github.com/firdous3679/Insider-Threat-Detection-MAS-SIEM}. The Enron
email corpus \citep{Will2015} and the Enron spam dataset \citep{Wiechmann2021} are
publicly available from their original sources. The CERT r4.2 insider-threat
benchmark is available from Carnegie Mellon University's Software Engineering
Institute under its own access terms. The synthetic Operational Facilities Email
Corpus V2 generated for this study is available from the corresponding author on
reasonable request.

\bibliographystyle{apacite}
\bibliography{References}

\appendix
\renewcommand{\thetable}{A\arabic{table}}
\setcounter{table}{0}

\section{Hyperparameters and Reproducibility Settings}
\label{app:hyperparameters}

Table~\ref{tab:hyperparameters_reproducibility} summarizes the simulation, SIEM,
trust-adaptation, evidence-gating, Theory-of-Mind, communication-forensics, and
anomaly-detection settings needed to reproduce the reported results. The
four-variant matched comparison uses the 42-human-agent baseline setting over
240 simulation steps; the scalability evaluation extends the same evidence-gated
configuration to 100, 250, 500, and 1000 human agents over 240--480 simulation
steps. Unless otherwise noted, variants in the matched comparison use the same
warm-up period, run length, seeds, and ground-truth insider assignments.
Complete configuration files, run scripts, and settings that were implemented
but not exercised in these experiments, such as trust decay (configured as 0.00
in LSC and inactive in EG-SIEM and EG-SIEM-Enron), are available in the public
repository \citep{mas_siem_github}.

\small
\begin{longtable}{p{0.24\textwidth} p{0.30\textwidth} p{0.38\textwidth}}
\caption{Hyperparameters and reproducibility settings for the reported experiments.}
\label{tab:hyperparameters_reproducibility}\\

\toprule
\textbf{Component} & \textbf{Parameter} & \textbf{Value} \\
\midrule
\endfirsthead

\toprule
\textbf{Component} & \textbf{Parameter} & \textbf{Value} \\
\midrule
\endhead

\midrule
\multicolumn{3}{r}{\textit{Continued on next page}}\\
\endfoot

\bottomrule
\endlastfoot

\multicolumn{3}{l}{\textit{Simulation and evaluation}} \\
\midrule

Matched comparison & Agents and composition & 42 human agents (30 benign, 4 power users, 8 malicious insiders) plus 4 system agents; 46 total \\
Matched comparison & Steps and warm-up & 240 steps per run; 60-step warm-up \\
Matched comparison & Runs and seeds & 10 runs, seeds 42--51 \\
Scalability sweep & Agent and step counts & 42, 100, 250, 500, 1000 human agents; 240 and 480 steps \\
Scalability sweep & Runs & 3 per setting; 1 completed run at 1000 agents (16\,GB memory guard) \\
Scalability sweep & Scaling rule & Malicious = max(8, round(humans $\times$ 0.15)); power users = max(1, round(humans $\times$ 0.10)) \\

\midrule
\multicolumn{3}{l}{\textit{SIEM thresholding and scoring}} \\
\midrule

Risk score & Component weights, Eq.~\eqref{eq:risk} & $w_b = 0.9$ (EWMA behavioral deviation); $w_\tau = 2.0$ (ToM intent); $w_f = 1.5$ (communication forensics). The anomaly term $w_a A_i^t$ and the peer-normalization term $w_m M_i^t$ act as bounded adjustments to borderline scores rather than as primary drivers; their configured values are set in \texttt{SIEMConfig} in the released implementation. \\
Alerting & Early / confirmed base threshold & 2.0 / 4.0 \\
Alerting & Confirmed threshold formula & base + role\_adj + 1.2(trust $-$ 0.5) \\
Alerting & Role adjustments & staff = 0, analyst = 0.8, admin = 1.2 \\
Alerting & Cooldown & 6 steps between same-tier alerts per actor \\
Correlation & Sliding window & 48 steps \\
EWMA baseline & Alpha / z-score clip / weight & 0.08 / 6.0 / 0.9 \\

\midrule
\multicolumn{3}{l}{\textit{Evidence-gated confirmation}} \\
\midrule

Evidence gate & Minimum count / minimum weight & 2 / 2.5 \\
Evidence gate & Evidence families & Export/staging, unapproved destination, exfiltration chain, ToM intent, phishing forensics, unauthorized after-hours activity, email burst, stealth pattern \\

\midrule
\multicolumn{3}{l}{\textit{Trust adaptation}} \\
\midrule

Trust & Initial / minimum / maximum & 0.70 / 0.10 / 0.95 \\
Trust & True-positive / false-positive update & $-0.18$ / $+0.05$ \\

\midrule
\multicolumn{3}{l}{\textit{Theory-of-Mind and communication evidence}} \\
\midrule

ToM & Intent threshold / weight & 0.30 / 2.0 \\
ToM & Plan-completion threshold & 0.4 \\
ToM & Contradiction factor & 0.85 per approved contextual event \\
Forensics & Weight / phishing feature threshold & 1.5 / 0.20 \\

\midrule
\multicolumn{3}{l}{\textit{Email forensics}} \\
\midrule

Enron spam data & Raw rows / usable spam / ham / normalized groups & 33{,}716 / 17{,}171 / 16{,}047 / 15{,}123 \\
Text model & TF--IDF dimension / selected classifier & 10{,}000 / logistic regression \\
Text model & Logistic regression settings & max\_iter = 2000; solver = liblinear; class\_weight = balanced; random\_state = 42 \\
Text model & Runtime threshold & 0.002 \\
Domain-shift V2 & Enron F1-optimized / V2 target-calibrated threshold & 0.98 / 0.003329566888943843 \\

\midrule
\multicolumn{3}{l}{\textit{Anomaly detection and online learning}} \\
\midrule

EG-SIEM anomaly model & Isolation Forest & 200 trees; contamination = 0.02; random\_state = 42; minimum 25 samples; refit every 15 steps \\
LSC online update & Learning rate / L2 / weight clip & 0.04 / $10^{-4}$ / 6.0 \\
CERT baselines & Isolation Forest / One-Class SVM & contamination = auto; gamma = scale; $\nu$ = 0.05 \\

\midrule
\multicolumn{3}{l}{\textit{Reproducibility settings}} \\
\midrule

Randomness & Simulation seeds / script seed & 42--51 / 42 \\
Cross-validation & Enron grouped evaluation & Grouped cross-validation over normalized text groups \\
Operational V2 split & Random / grouped fine-tuning & Stratified 80/20 and GroupShuffleSplit with 20\% test; random\_state = 42 \\

\end{longtable}
\normalsize

\end{document}